\documentclass[titlepage,10pt]{article}

\usepackage[scaled=0.97]{helvet}
\usepackage[T1]{fontenc}

\usepackage{array}
\usepackage{amsmath,amsbsy,amsthm}
\usepackage{amsxtra}
\usepackage{amstext}
\usepackage{amssymb}
\usepackage{latexsym}

\usepackage{url}
\usepackage{graphicx}

\usepackage[usenames]{color}   

\usepackage{setspace}          
\usepackage{natbib}
\bibpunct{}{}{,}{s}{}{;}

\usepackage{booktabs}
\usepackage{geometry}

\usepackage{titlesec}          
\usepackage{enumitem}          
\usepackage{wrapfig}           

\def\E{\mathop{\rm E\,\!}\nolimits}

\newcommand{\ba}{\boldsymbol{a}}

\newcommand{\br}{\boldsymbol{r}}

\newcommand{\by}{\boldsymbol{y}}

\newcommand{\bA}{\boldsymbol{A}}

\newcommand{\bH}{\boldsymbol{H}}
\newcommand{\bI}{\boldsymbol{I}}

\newcommand{\bN}{\boldsymbol{N}}

\newcommand{\bQ}{\boldsymbol{Q}}
\newcommand{\bR}{\boldsymbol{R}}

\newcommand{\bW}{\boldsymbol{W}}

\newcommand{\bbeta}{\boldsymbol{\beta}}

\newcommand{\bnu}{\boldsymbol{\nu}}

\newcommand{\bsigma}{\boldsymbol{\sigma}}
\newcommand{\btheta}{\boldsymbol{\theta}}

\newcommand{\bDelta}{\boldsymbol{\Delta}}

\newcommand{\bOmega}{\boldsymbol{\Omega}}
\newcommand{\bPhi}{\boldsymbol{\Phi}}

\newcommand{\Mendel}{{\sc Mendel}}
\newcommand{\fastlmm}{{\sc FaST-LMM}}
\newcommand{\gemma}{{\sc GEMMA}}

\title{Fast Genome-Wide QTL Association Mapping on Pedigree and Population Data}
\author{
Hua Zhou,
\thanks{Department of Statistics,
North Carolina State University, 
Raleigh, NC 27695-8203} 
\and John Blangero,
\thanks{Department of Genetics,
Texas Biomedical Research Institute,
San Antonio, TX 78245-0549}
\and Thomas D. Dyer,$^{2}$
\and Kei-hang K. Chan,
\thanks{Department of Human Genetics,
University of California,
Los Angeles, CA 90095-1766}
\thanks{Department of Epidemiology,
University of California,
Los Angeles, CA 90095-1766}
\and Kenneth Lange,$^{3}$
\thanks{Department of Biomathematics,
University of California,
Los Angeles, CA 90095-1766}
\thanks{Department of Statistics,
University of California,
Los Angeles, CA 90095-1766}
\and Eric M. Sobel$^{3}$
}

\begin{document}
\baselineskip=20pt
\maketitle

\doublespacing
\begin{abstract}

Since most analysis software for genome-wide association studies (GWAS) currently exploit only unrelated individuals, there is a need for efficient applications that can handle general pedigree data or mixtures of both population and pedigree data. Even data sets thought to consist of only unrelated individuals may include cryptic relationships that can lead to false positives if not discovered and controlled for. In addition, family designs possess compelling advantages. They are better equipped to detect rare variants, control for population stratification, and facilitate the study of parent-of-origin effects. Pedigrees selected for extreme trait values often segregate a single gene with strong effect. Finally, many pedigrees are available as an important legacy from the era of linkage analysis. Unfortunately, pedigree likelihoods are notoriously hard to compute. In this paper we re-examine the computational bottlenecks and implement ultra-fast pedigree-based GWAS analysis. Kinship coefficients can either be based on explicitly provided pedigrees or automatically estimated from dense markers. Our strategy (a) works for random sample data, pedigree data, or a mix of both; (b) entails no loss of power; (c) allows for any number of covariate adjustments, including correction for population stratification; (d) allows for testing SNPs under additive, dominant, and recessive models; and (e) accommodates both univariate and multivariate quantitative traits. On a typical personal computer (6 CPU cores at 2.67 GHz), analyzing a univariate HDL (high-density lipoprotein) trait from the San Antonio Family Heart Study (935,392 SNPs on 1357 individuals in 124 pedigrees) takes less than 2 minutes and 1.5 GB of memory. Complete multivariate QTL analysis of the three time-points of the longitudinal HDL multivariate trait takes less than 5 minutes and 1.5 GB of memory. The algorithm is implemented as the Ped-GWAS Analysis (Option 29) in the \Mendel\ statistical genetics software package, which is freely available for Macintosh, Linux, and Windows platforms from \url{http://genetics.ucla.edu/software/mendel}.

{\bf Key words}: genome-wide association study; pedigree; kinship; score test; fixed-effects models, multivariate traits 
\end{abstract}

\section{Introduction}

Genome-wide association studies (GWAS) are now at a crossroads. After the discovery of thousands of genes influencing hundreds of common traits \citep{Hindorff2009}, much of the low-hanging fruit has been plucked \citep{Ku2010,Visscher2012}.  Because of the enormous sample sizes of current studies, new trait genes are still being uncovered. Unfortunately, most entail small effects.  Is it possible that inheritance is predominantly polygenic, and a law of diminishing returns has set in?  The push to exploit rare variants is one response to this dilemma. The previous generation of geneticists relied on linkage to map rare variants. Linkage mapping fell from grace because of its poor resolution.  Reducing a genome search to a one or two megabase region leaves too large an expanse of DNA to sift. The real gold of linkage mapping may well be its legacy pedigrees \citep{Ott2011}. Pedigree data is particularly attractive in association studies because it permits control of population substructure and study of parent-of-origin effects. Related affecteds are also more likely to share the same disease predisposing gene than unrelated affecteds. Even in population-based association studies, taking into account estimated identity-by-descent (IBD) information is apt to reduce false positives and increases power. The recent availability of dense marker data from genotyping chips enables quick and accurate estimation of global and even local IBD \citep{Day-Williams2011}.

\begin{table*}[t]
\centering
\begin{tabular}{lccc}
\toprule
                                           & \Mendel & \fastlmm & \gemma \\
\midrule\midrule
Multi-threaded operation                   & Yes     & Yes      & No     \\
Can estimate kinships via SNPs             & Yes     & Yes      & Yes    \\
Imports \& exports kinship estimates       & Yes     & Yes      & Yes    \\
Allows retained co-variates                & Yes     & Yes      & Yes    \\
Allows linear constraints on co-variates   & Yes     & No       & No     \\
Can use either LRT or score test           & Yes     & No       & Yes*   \\
Allows multivariate analysis               & Yes     & No       & Yes    \\
Can perform multiple univariate analyses   & Yes     & No       & No     \\
Allows $> 2$ variance components           & Yes     & No       & No     \\
Analyzes X-linked loci                     & Yes     & No       & No     \\
Automatic SNP filtering on MAF             & Yes     & No       & Yes    \\
Allows non-additive SNP models             & Yes     & No       & No     \\
Detects outlier pedigrees                  & Yes     & No       & No     \\
Detects outlier individuals                & Yes     & No       & No     \\
Can simulate genotype/phenotype data       & Yes     & No       & No     \\
Reads in fractional genotype values        & No      & Yes      & Yes    \\
\bottomrule \\
\end{tabular}
\caption{Comparison of features in \Mendel, \fastlmm, and \gemma\ for GWAS of QTLs. *\gemma\ can use the likelihood ratio, score, or Wald test.}
\label{table:software-comparison}
\vspace{-0.25in}
\end{table*}

Geneticists turned to random sample and case-control data because of the relative ease of collecting population data and the computational challenges posed by pedigrees. The tide of computational complexity is now beginning to turn.  To handle pedigree data in association testing, statistical geneticists have proposed semiparametric methods such as the generalized linear mixed model (GLMM) \citep{Amin07,Aulchenko07} and generalized estimating equations (GEE) \citep{ChenYang09GWAF,ChenLiuWei11GWAFComparison}. Although such methods work for both quantitative and binary traits, they are compromised by current restrictions that reduce power. The GEE approach requires input of a working correlation structure for each pedigree. The kinship coefficient matrix is a natural candidate. However, current implementations require the same working correlation matrix across all clusters, which implicitly requires all pedigrees to have the same structure \citep{ChenLiuWei11GWAFComparison}. This is a dubious and restrictive assumption. In the limited context of case-control studies, recent methods such as $M_{QLS}$ \citep{MQLS2007}, {\sc roadtrips} \citep{Thornton10Roadtrips}, and {\sc FPCA} \citep{ZhuXiong2012} correct for pedigree and ethnically induced correlations by exploiting dense marker data. Other authors attack the same issues more broadly from the GLMM perspective \citep{EMMAX2010,TASSEL2010,FaST2011}. \citet{MTMM2012} generalizes GLMM to multivariate traits. Models based on the transmission-disequilibrium test (TDT) \citep{SpielmanEwens98TDT} and its generalization, the family-based association test (FBAT) \citep{LairdHorvathXu2000FBAT,LangeLaird02FBAT,PBAT2005,PBAT2009-Stats,PBAT2009-PLoS}, are promising but ignore covariates and polygenic background. See \citet{VanSteen2011} for a recent overview of FBAT methods for GWAS. We treat all of these extensions in a unified framework consistent with exceptionally fast computing. 

The present paper re-examines the computational bottlenecks encountered in association mapping with pedigree data.  It turns out that the previous objections to pedigree GWAS can be overcome. Kinship coefficients can be based on explicitly provided pedigree structure or estimated from dense markers when genealogies are missing or dubious. Frequentist hypothesis testing usually operates by comparing maximum likelihoods under the null and alternative hypotheses.  Maximization of the alternative likelihood must be conducted for each and every marker. Score tests constitute a more efficient strategy than likelihood ratio tests. This is the point of departure taken by \citet{Chen2007}, but they use approximations that we avoid. The {\sc glogs} program \citep{GLOGS2012} makes similar approximations in the case-control setting. Here we consider arbitrary pedigrees and multivariate quantitative traits. Score tests require no additional iteration under the alternative model.  All that is needed is evaluation of a quadratic form combining the score vector and the expected information matrix at the maximum likelihood estimates under the null model. Although it takes work to assemble these quantities, a careful analysis of the algorithm shows that fast testing is perfectly feasible.

In our implementation of score testing, the few SNPs with the most significant score-test p-values are automatically re-analyzed by the slightly more powerful, but much slower, likelihood ratio test (LRT). Our fixed effects (mean component) model assumes Gaussian variation of the trait; the two alleles of a SNP shift trait means.  There is no confounding of association and linkage. This framework carries with it several advantages. First, it applies to random sample data, pedigree data, or a mix of both. Second, it enables covariate adjustment, including correction for population stratification. Third, it accommodates additive, dominant, and recessive SNP models. Fourth, it also accommodates both univariate and multivariate traits. And fifth, as just mentioned, it fosters both likelihood ratio tests and score tests.  The mean component model is now implemented in our software package \Mendel\ for easy use by the genetics community. In addition, \Mendel\ provides a complete suite of tools for pedigree analysis, including GWAS data preparation and manipulation, pedigree genotype simulation (gene dropping), trait simulation, genotype imputation, local and global kinship coefficient estimation, and pedigree-based GWAS (ped-GWAS) \citep{Lange2005Association,Lange2013Mendel}.  
 
The competing software packages {\sc EMMAX} \citep{Kang2008}, {\sc MMM} \citep{PirinenDonnellySpencer2013MMM}, \fastlmm\ \citep{FaST2011,FaST2012}, and \gemma\ \citep{ZhouStephens12GEMMA,ZhouStephens14GEMMA} already implement variance component models for quantitative trait locus (QTL) analysis. Exhaustive comparison of \Mendel\ to each of these programs is beyond the scope of the current paper. We limit our comparisons of \Mendel\ to the state-of-art packages \fastlmm\ and \gemma, arguably the fastest and most sophisticated of the competition. Table~\ref{table:software-comparison} summarizes some of the qualitative features of these packages. Our numerical examples also demonstrate an order of magnitude advantage in speed of \Mendel\ over \fastlmm\ and \gemma. This advantage stems from our careful formulation of the score test and our exploitation of the multicore processors resident in almost all personal computers and computational clusters.

\section{Methods}

\subsection{QTL Association Mapping with Pedigrees}

QTL association mapping typically invokes the multivariate Gaussian distribution to model the trait values $\by=(y_{i})$ over a pedigree. The observed trait value $y_i$ of person $i$ can be either univariate or multivariate.  For simplicity we first assume $y_i$ is univariate and later indicate the necessary changes for multivariate $y_i$. The standard model \citep{Lange02GeneticsBook}
collects the corresponding trait means into a vector $\bnu$ and the corresponding covariances into a matrix $\bOmega$ and represents the loglikelihood of a pedigree as
\begin{eqnarray}
L & = & -\frac{1}{2} \ln \det \bOmega
        -\frac{1}{2} (\by-\bnu)^t \bOmega^{-1}(\by-\bnu), \label{eqn:normal_loglikelihood}
\end{eqnarray}
where $\det$ denotes the determinant function and the covariance matrix is typically parametrized as
\begin{eqnarray}
	\bOmega & = & 2 \sigma_a^2 \bPhi + \sigma_d^2 \bDelta_7 + \sigma_h^2 \bH + \sigma_e^2 \bI.
	\label{variance_decomposition}
\end{eqnarray}
Here the variance component $\bPhi$ is the global kinship coefficient matrix capturing additive polygenic effects, and $\bDelta_7$ is a condensed identity coefficient matrix capturing dominance genetic effects. When pedigree structure is explicitly given, these genetic identity coefficients are easily calculated \citep{Lange02GeneticsBook}. With unknown or dubious genealogies, the global kinship coefficient can be accurately estimated from dense markers \citep{Day-Williams2011}. The household effect matrix $\bH$ has entries $h_{ij}=1$ if individuals $i$ and $j$ belong to the same household and 0 otherwise. Individual environmental contributions and trait measurement errors are incorporated via the identity matrix $\bI$. \Mendel's implementation of this model can include both the two standard variance classes, additive and environmental, as well as the two extra variances classes, dominance and household. Inclusion of additional variance classes has no significant effect on \Mendel's speed of computation.

\begin{table}[t]
\centering
\begin{tabular}{c||ccc}
\toprule
{\bf Genotype} & {\bf Additive} & {\bf Dominant} & {\bf Recessive} \\
\midrule\midrule
\verb|1/1|     &       --1      &      --1       &       --1       \\
\verb|1/2|     &         0      &      --1       &        +1       \\
\verb|2/2|     &        +1      &       +1       &        +1       \\
\bottomrule
\end{tabular}
\vspace{0.05in}
\caption{Genotype encodings for the major gene models. The additive model is the default choice. In the genotype column, ``1'' and ``2'' represent the first and second alleles for each SNP. An effect size estimate reflects the change in trait values due to each positive unit change in the encodings. For example, the default additive model estimates the mean trait difference in moving from a 1/2 genotype to a 2/2 genotype.}
\label{table:genotype-code}
\end{table}

In general, a mixed model for QTL association mapping captures polygenic and other random effects through $\bOmega$ and captures QTL fixed effects through $\bnu$. Let $\bbeta$ denote the full vector of regression coefficients parameterizing $\bnu$.  In a linear model one postulates that $\bnu = \bA \bbeta$ for some predictor matrix $\bA$ incorporating relevant covariates such as age, gender, and diet. In testing association against a given SNP, $\bA$ is augmented by an extra column whose entries encode genotypes according to one of the models (additive, dominant, and recessive) shown in see Table~\ref{table:genotype-code}. To accommodate imprecise imputation in an additive model, these encodings can be made fractional. The corresponding component of $\bbeta$, $\beta_\text{SNP}$, is the SNP effect size. In likelihood ratio association testing one contrasts the null hypothesis $\beta_\text{SNP} = 0$ with the alternative hypothesis $\beta_\text{SNP} \ne 0$. In testing a univariate trait, the likelihood ratio statistic asymptotically follows a $\chi_1^2$ distribution. In testing a multivariate trait with $T>1$ components, each row of $\bA$ must be replicated $T$ times. The likelihood ratio statistic then asymptotically follows a $\chi_T^2$ distribution. To implement likelihood ratio testing, iterative maximum likelihood estimation must be undertaken for each and every SNP under the alternative hypothesis. This unfortunate requirement is the major stumbling block retarding pedigree analysis.

Score tests serve as convenient substitutes for likelihood ratio tests. The current paper describes how to implement ultra-fast score tests for screening SNPs. Only SNPs with the most significant score test p-values are further subjected to the more accurate likelihood ratio test. An advantage of the likelihood ratio method is that it estimates effect sizes.  In contrast, the score test only requires parameter estimates under the null hypothesis and involves no iteration beyond fitting the null model.  The score vector is the gradient $\nabla L(\btheta)$ of the loglikelihood $L(\btheta)$, where the full parameter vector $\btheta$ includes variance components such as the additive genetic variance in addition to the regression coefficient vector $\bbeta$.  The transpose $dL(\btheta)$ of the score is a row vector called the first differential of $L(\btheta)$. The expected information $J(\btheta)$ is the covariance matrix of the score vector. It is well known that the expected value of the observed information matrix (negative second differential) $-d^2 L(\btheta)$ coincides with $J(\btheta)$ \citep{Rao2009}.  The score statistic
\begin{eqnarray*}
S(\btheta) & = & dL(\btheta) J(\btheta)^{-1} \nabla L(\btheta) \;\;\, \approx \;\;\,
dL(\btheta) [-d^2L(\btheta)]^{-1} \nabla L(\btheta)
\end{eqnarray*}
is evaluated at the maximum likelihood estimates under the null hypothesis with the parameter $\beta_\text{SNP}$ of the alternative hypothesis set to 0.

\subsection{Fast Score Test for Individual SNPs}

Under the multivariate model, the expected information matrix $J(\btheta)$ for a single pedigree can be written in the block diagonal form
\begin{eqnarray}
J(\btheta) & = & \begin{pmatrix}
\E[- d_{\bbeta}^2 L(\btheta)] & 0 \cr 0 & \E[ - d_{\bsigma}^2 L(\btheta)]
\end{pmatrix},  \label{eqn:expected-information}
\end{eqnarray}
where $\bsigma$ denotes the vector of variance parameters \citep{Lange02GeneticsBook}. For independent pedigrees, the loglikelihoods (\ref{eqn:normal_loglikelihood}) and corresponding score vectors and expected information matrices add.  Hence, the block diagonal form of $J(\btheta)$ is preserved.  Because the inverse of a block diagonal matrix is block diagonal, the score statistic splits into a piece contributed by the variance components plus a piece contributed by the mean components. The maximum likelihood estimate $\hat{\btheta}=(\hat{\bbeta},\hat{\bsigma})$ under the null model is a stationary point of the loglikelihood. Thus, the variance components segment $\nabla_{\bsigma} L (\hat{\btheta})$ of the score vector vanishes.  We therefore focus on the mean components segment of the score vector.

If the pedigrees are labeled $1,\ldots,n$, then the pertinent quantities for implementing the score test are
\begin{eqnarray*}
\sum_{i=1}^n \nabla_{\bbeta} L_i (\btheta)   & = & \sum_{i=1}^n \bA_i^t \bOmega_i^{-1} \br_i  \\
\sum_{i=1}^n \E[- d_{\bbeta}^2 L_i(\btheta)] & = & \sum_{i=1}^n \bA_i^t \bOmega_i^{-1} \bA_i ,
\end{eqnarray*}
where $\br_i = \by_i - \bA_i \hat{\bbeta}$ is the residual for pedigree $i$ and the covariance matrix $\bOmega_i$ for pedigree $i$ is determined by equation (\ref{variance_decomposition}). See Chapter 8 of \citet{Lange02GeneticsBook} for a detailed derivation of the score and expected information. Since the score statistic is calculated from estimated parameters under the null model, residuals do not change when we expand the null model to the alternative model keeping $\beta_\text{SNP}=0$. Calculation of the maximum likelihood estimate $\hat{\btheta}$ under the null is accomplished by a
quasi-Newton algorithm whose initial step reduces to Fisher scoring \citep{Lange1976,Lange02GeneticsBook}.

For pedigree $i$ under the alternative hypothesis, the design matrix $\bA_i$ can be written as $( \ba_i,\bN_i)$, where $\bN_i$ is the design matrix under the null hypothesis and $\ba_i$ conveys the genotypes at the current SNP. In testing a univariate trait, the entries of $\ba_i$ are taken from Table~\ref{table:genotype-code}. If allele counts are imputed under the additive model, then the entries of $\ba_i$ may be fractional numbers drawn from the interval $[-1,1]$. In testing a multivariate trait with $T>1$ components, each row of $\bA_i =(\ba_i,\bN_i)$ must be replicated $T$ times. The only exceptions to this rule occur for people missing some but not all component traits;
otherwise, the covariance matrix $\bOmega_i$ for pedigree $i$ decomposes into a sum of Kronecker products \citep{Lange02GeneticsBook}. Regardless of whether the trait is univariate or multivariate, one must compute the quantities
\begin{eqnarray*}
\sum_{i=1}^n \nabla_{\bbeta} L_i (\btheta)   & = & \begin{pmatrix}
\sum_{i=1}^n \ba_i^t \bOmega_i^{-1} \br_i \\
\sum_{i=1}^n \bN_i^t \bOmega_i^{-1} \br_i
\end{pmatrix} \\
\sum_{i=1}^n \E[- d_{\bbeta}^2 L_i(\btheta)] & = & \begin{pmatrix}
 \sum_{i=1}^n \ba_i^t \bOmega_i^{-1} \ba_i  & \sum_{i=1}^n \ba_i^t \bOmega_i^{-1} \bN_i\\
\sum_{i=1}^n \bN_i^t \bOmega_i^{-1} \ba_i  & \sum_{i=1}^n \bN_i^t \bOmega_i^{-1} \bN_i
\end{pmatrix}.
\end{eqnarray*}
At the maximum likelihood estimates under the null model, the partial score vector $\sum_{i=1}^n \bN_i^t \bOmega_i^{-1} \br_i$ vanishes. Hence, the score statistic for testing a SNP can be expressed as
\begin{eqnarray*}
 S & \!\!\! = \!\!\! & \bR^t \left[ \bQ -  \bW^t \left( \sum_{i=1}^n \bN_i^t \bOmega_i^{-1} \bN_i \right)^{-1} \bW \right]^{-1} \bR,
\end{eqnarray*}
where
\begin{eqnarray*}
           \bQ   &  =  &   \sum_{i=1}^n \ba_i^t \bOmega_i^{-1} \ba_i, 
\quad \bR \;\; = \;\; \sum_{i=1}^n \ba_i^t \bOmega_i^{-1} \br_i, \\
\quad \bW & = & \sum_{i=1}^n \bN_i^t \bOmega_i^{-1} \ba_i.
\end{eqnarray*}
In the score statistic $S$, the covariance matrices $\bOmega_i^{-1}$ and residual vectors $\br_i$ are evaluated at the maximum likelihood estimates under the null model. Large sample theory says that $S$ asymptotically follows a $\chi_T^2$ distribution.

These formulas suggest that we precompute and store the quantities $ \bOmega_i^{-1}$, $\bOmega_i^{-1}\bN_i$, and $\bOmega_i^{-1}\br_i$ for each pedigree $i$ and the overall sum $\sum_{i=1}^n \bN_i^t \bOmega_i^{-1} \bN_i$ at the maximum likelihood estimates under the null hypothesis. From these parts, the basic elements of the score statistic can be quickly assembled.  The most onerous quantity that must be computed on the fly as each new SNP is encountered is $\sum_{i=1}^n \ba_i^t \bOmega_i^{-1} \ba_i $.  If there are $p_i$ people in pedigree $i$, then computation of the quadratic form $\ba_i^t \bOmega_i^{-1} \ba_i $ requires $O(p_i^2)$ arithmetic operations.  This looks worse than it is in practice since the entries of $\ba_i$ are integers (--1, 0, and 1) in the absence of fractional imputation.  This simplification allows one to avoid a fair amount of arithmetic. Assembling the remaining parts of the score statistic requires $O(p_i)$ arithmetic operations.

Individuals missing univariate trait values are omitted from analysis. Individuals missing some but not all components of a multivariate trait are retained in analysis. The proper adjustments for missing data are made automatically in the score statistic because sections of Gaussian random vectors are Gaussian.

SNPs with minor allele counts below a user-designated threshold are also omitted from analysis. Note that if the minor allele count across a study is 0, then the given SNP is mono-allelic and worthless in association testing. \Mendel's default threshold of 3 is motivated by the rule of thumb in contingency table testing that all cells have an expected count of at least 3. For a multivariate trait, a SNP may fall below the threshold for some component traits but not for others. This situation can occur when each trait displays a different pattern of missing data across individuals. \Mendel\ retains such anomalous SNPs only for those component traits with a sufficient number of minor alleles. Again, proper adjustments are made automatically within the score test statistic to account for partial data.

\Mendel's analysis yields a score test p-value for each SNP. For the user-designated most significant SNPs, \Mendel's subsequent likelihood ratio test outputs an estimated SNP effect size, a standard error of that estimate, and the fraction of the total variance explained by that SNP. For a multivariate trait, \Mendel\ outputs a SNP effect size and associated standard error for each component trait. In the initial analysis under the null model with no SNPs, \Mendel\ provides estimates with standard errors of all mean and variance components included in the model. Finally, an estimate of heritability with standard error is also provided.

The extension of the score test to the multivariate $t$-distribution is straightforward \citep{Lange89Robust}. Suppose $\eta$ equals the degrees of freedom of the $t$-distribution and $m_i$ equals the number of observed person-trait combinations for pedigree $i$.  The sections of the score and expected information pertinent to the mean components for the pedigree reduce to 
\begin{eqnarray*}
\nabla_{\bbeta} L_i (\btheta)   & = & \frac{\eta+m_i}{\eta+s_i} \bA_i^t \bOmega_i^{-1} \br_i  \\
\E[- d_{\bbeta}^2 L_i(\btheta)] & = & \frac{\eta+m_i}{\eta+m_i+2} \bA_i^t \bOmega_i^{-1} \bA_i ,
\end{eqnarray*}
where $r_i$ is the residual and $s_i = \br_i^t \bOmega_i^{-1} \br_i$ is the associated Mahalanobis distance. A sensible choice for $\eta$ is its estimate under the null model. 

\subsection{Kinship Estimation From SNPs}

\Mendel\ can either calculate the global kinship coefficient matrix $\bPhi$ from the provided pedigree structures or estimate it from dense genotypes. In global kinship estimation \Mendel's default uses an evenly spaced 20\% of the available SNPs, and only compares pairs of individuals within defined pedigrees. Hence, $\bPhi$ is block diagonal. Users can trivially elect to exploit a larger fraction of the available SNPs or estimate kinship for {\em all} pairs of individuals. Given $S$ selected SNPs, \Mendel\ estimates the global kinship coefficient of individuals $i$ and $j$ based on either the genetic relation matrix (GRM) method 
\begin{eqnarray*}
\hat \Phi_{ij} & = & \frac{1}{2S} \sum_{k=1}^S \frac{(x_{ik} - 2p_k)(x_{jk} - 2p_k)}{2p_k(1-p_k)}
\end{eqnarray*}
or the method of moments (MoM) \citep{Day-Williams2011,Lange14NextGenStatGene}
\begin{eqnarray*}
\hat \Phi_{ij} & = & \frac{e_{ij} - \sum_{k=1}^S \left[p_k^2 + (1-p_k)^2 \right]}{S - \sum_{k=1}^S\left[p_k^2 + (1-p_k)^2\right]},
\end{eqnarray*}
where $p_k$ is the minor allele frequency at SNP $k$, $x_{ik}$ is the number of minor alleles in $i$'s genotype at SNP $k$, and
\begin{eqnarray*}
	e_{ij} & = & \frac{1}{4} \sum_{k=1}^S \left[ x_{ik}x_{jk} + (2-x_{ik})(2-x_{jk}) \right]
\end{eqnarray*}
is the observed fraction of alleles identical by state (IBS) between $i$ and $j$. The GRM method is \Mendel's default. In general, one can think of the GRM method centering and scaling each genotype, while the MoM method uses the raw genotypes and then centers and scales the final result.

\subsection{Other Utilities for Handling Pedigree Data}

To encourage thorough testing of new statistical methods, such as the current Ped-GWAS score test, we have implemented both genotype and trait simulation in our genetic analysis program \Mendel\ \citep{Lange2013Mendel}. \Mendel\ does genotype simulation (gene dropping) subject to prescribed allele frequencies, a given genetic map, and Hardy-Weinberg and linkage equilibrium.  If one fixes founder haplotypes and simulates conditional on these, then the unrealistic assumption of linkage equilibrium can be relaxed. Missing data patterns are respected or imposed by the user. It is also possible to set the rate for randomly deleting data and to simulate genotypes for people of mixed ethnicity by defining different ancestral populations, each with its own allele frequencies. If this feature is invoked, then each pedigree founder should be assigned to a population.

Trait simulation can be layered on top of genotype simulation. \Mendel\ simulates either univariate traits determined by generalized linear models or multivariate Gaussian traits determined by variance component models. The biggest limitations are the restriction to a single major locus and the generalized linear model assumption that trait correlations are driven solely by this locus.  Variance component models enable inclusion of environmental effects and more complicated correlations among relatives. In the variance component setting, univariate as well as multivariate Gaussian traits can be simulated. Most variance component models are built on Gaussian distributions, but \Mendel\ allows one to replace these by multivariate $t$-distributions. Thus, users can investigate robust statistics less prone to distortion by outliers. More theoretical and implementation details appear in the \Mendel\ documentation \citep{Lange2013Mendel}.

\section{Results}

\subsection{The San Antonio Family Heart Study}

We analyzed a real data set collected by the San Antonio Family Heart Study (SAFHS) \citep{SAFHS1996}. The data consist of 3637 individuals in 211 Mexican American families. High-density lipoprotein (HDL) levels were measured at up to three time points for each of the 1429 phenotyped individuals. These traits are denoted HDL$_1$, HDL$_2$, and HDL$_3$, measured at corresponding ages AGE$_1$, AGE$_2$, and AGE$_3$. Some of the phenotyped individuals have HDL measurements at only one or two of the time points.  Of the 1429 phenotyped individuals, 1413 were genotyped at 944,427 genome-wide SNPs. The genotyping success rate exceeded 98\% in 1388 of these individuals over 124 pedigrees. The largest family contains 247 individuals; five others also contain more than 90 individuals. The smallest pedigree was a singleton. Genotyping success rates were above 98\% for 935,392 SNPs.

\subsection{Comparison with \fastlmm\ and \gemma}

\begin{table*}[t]
\vspace{-0.10in}
\centering
\begin{tabular}{l||lrrlrrrr}
\toprule
Trait   &    SNP     & Chr. &  Base Pair  &   MAF    & $-\log_{10}(\text{p-val})$ & $-\log_{10}(\text{p-val})$ & $-\log_{10}(\text{p-val})$ & $-\log_{10}(\text{p-val})$ \\
        &            &      &  Position   &          & \Mendel\ default           & \Mendel\ all-pairs         & \fastlmm                   & \gemma    \\
\midrule\midrule
        & rs7303112  &  12  &  97,596,023 & 0.00455  &                10.21       &                10.71       &   7.63                     &  7.24     \\
HDL$_1$ & rs8040647  &  15  &  32,304,988 & 0.00147* &                 7.44       &                 7.56       &   7.35                     &  7.45     \\
        & rs9972594  &  15  &  32,421,102 & 0.00147* &                 7.44       &                 7.56       &   7.37                     &  7.46     \\
        & rs7167103  &  15  &  32,830,477 & 0.00147* &                 7.44       &                 7.56       &   7.35                     &  7.44     \\
\midrule
HDL$_2$ & rs7100957  &  10  &  28,207,332 & 0.00183* &                 8.84       &                 8.95       &   8.88                     &  8.82     \\
\midrule
HDL$_3$ & rs17060933 &   8  &  22,510,029 & 0.00382  &                 8.23       &                 8.28       &   8.61                     &  8.59     \\
\midrule
        & rs7303112  &  12  &  97,596,023 & 0.00644  &                 9.89       &                 9.94       &                            &           \\
HDL$_\text{Joint}$
        & rs16925210 &  10  &  25,308,103 & 0.00217  &                 8.15       &                 8.33       &                            &           \\
{\small with}
        & rs7091416  &  10  &  25,318,381 & 0.00217  &                 8.15       &                 8.33       &       Not                  &       Not \\
{\small constrained}
        & rs10075658 &   5  & 148,911,957 & 0.00144* &                 8.16       &                 8.21       & Available                  & Available \\
{\small covariates}
        & rs7733139  &   5  & 145,977,990 & 0.00217  &                 7.36       &                 7.34       &                            &           \\
        & rs7100957  &  10  &  28,207,332 & 0.00870  &                 7.20       &                 7.30       &                            &           \\
\midrule
        & rs7303112  &  12  &  97,596,023 & 0.00644  &                 9.82       &                 9.88       &                            & 11.08     \\
HDL$_\text{Joint}$
        & rs16925210 &  10  &  25,308,103 & 0.00217  &                 8.04       &                 8.23       &                            &  3.53     \\
{\small without}
        & rs7091416  &  10  &  25,318,381 & 0.00217  &                 8.04       &                 8.23       &       Not                  &  3.52     \\
{\small constrained}
        & rs10075658 &   5  & 148,911,957 & 0.00144* &                 8.12       &                 8.17       & Available                  &  3.47     \\
{\small covariates}
        & rs7733139  &   5  & 145,977,990 & 0.00217  &                 7.41       &                 7.40       &                            &  3.47     \\
        & rs7100957  &  10  &  28,207,332 & 0.00870  &                 7.19       &                 7.30       &                            &  4.48     \\
        & rs10083226 &  13  & 104,434,452 & 0.00219  &                 7.10       &                 7.31       &                            &  2.14     \\
\bottomrule
\end{tabular}
\vspace{0.05in}
\caption{All SNPs with minor allele frequency (MAF) above 0.001 that reach genome-wide significance in any of the analyses of the HDL traits from the San Antonio Family Heart Study (SAFHS). All default parameters were used except for minor changes to the quality control thresholds (see text). Also, \Mendel\ was run in both default and all-pairs modes. \Mendel's default mode estimates non-zero global kinship coefficients only for pairs of individuals within the same input pedigree; \Mendel\ in all-pairs mode, \fastlmm, and \gemma\ estimate coefficients for all pairs of individuals. Genome-wide significance was declared for p-values $< 5\times 10^{-8} \implies -\log_{10}(\text{p-value}) > 7.3$. The SAFHS has 1413 genotyped and phenotyped individuals in 124 pedigrees. The genotypes include roughly 1 million SNPs. The phenotypes include the subjects' high-density lipoprotein (HDL) level and age at three time points. The HDL$_\text{Joint}$ runs are multivariate analyses of HDL$_1$, HDL$_2$, and HDL$_3$ jointly; all other runs are univariate analyses. See the text for a list of the covariates used in each analysis. Note that in the multivariate analysis, \Mendel\ is able to use roughly twice as many individuals as \gemma\ (see text and Table~\ref{table:run-time}), which may explain the less significant findings for \gemma. Each MAF is based on the pedigree founders, except where marked by an asterisk (*). In these cases the minor allele did not appear in the genotyped founders, and its frequency was estimated from all genotyped individuals.}\label{table:HDL}
\vspace{-0.25in}
\end{table*}

For fair comparisons, we directed \Mendel\ to estimate SNP-based global kinship coefficients for all pairs of individuals ignoring the input pedigrees. This is the default in \fastlmm\ and \gemma. In addition, we ran \Mendel's default in which the coefficients are estimated only for pairs of individuals within the same input pedigree. We also slightly adjusted some of the default quality control thresholds so the programs would be analyzing roughly the same set of SNPs and individuals. For example, by default \Mendel\ filters SNPs with fewer than three occurrences of the minor allele in the data; in contrast, \fastlmm\ only filters SNPs with zero occurrences of the minor allele, and \gemma\ filters SNPs with minor allele frequency (MAF) $< 0.01$. All other defaults were observed throughout. Users can easily adjust the \Mendel\ analysis parameters via its control file and the \fastlmm\ and \gemma\ analysis parameters via their command line. 

We first carried out three univariate QTL analyses of HDL$_1$, HDL$_2$, and HDL$_3$, using SEX and AGE$_1$, AGE$_2$, or AGE$_3$ as covariates. We then ran a multivariate QTL analysis of HDL$_1$, HDL$_2$, and HDL$_3$ jointly, which we refer to as HDL$_\text{Joint}$. For the multivariate analysis, the most appropriate configuration is to constrain the effects of the SEX and AGE covariates to be the same on all three HDL measurements. Such linear constraints are imposed in \Mendel\ via a few simple lines in its control file. \fastlmm\ and \gemma\ do not allow constraints on covariates. Therefore, we also ran a multivariate analysis with only the SEX covariate and no constraints. With no constraints, SEX will have a slightly different effect on each component phenotype in the multivariate analysis. For instance, \Mendel's default run estimated a female effect of $2.5 \pm 0.3$ on HDL$_1$, $2.1 \pm 0.4$ on HDL$_2$, and $2.7 \pm 0.4$ on HDL$_3$. \fastlmm\ cannot perform any multivariate analyses.

\begin{figure}[tb]
\vspace{-0.10in}
\centering
\includegraphics[width=3.2in]{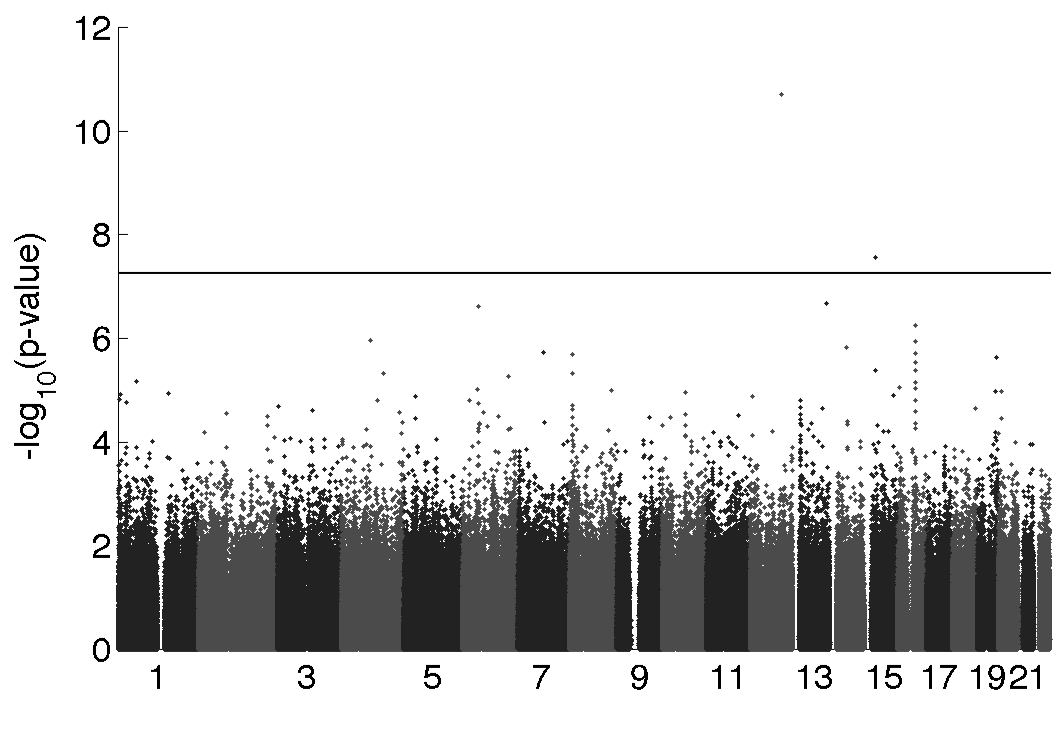}
\includegraphics[width=3.0in]{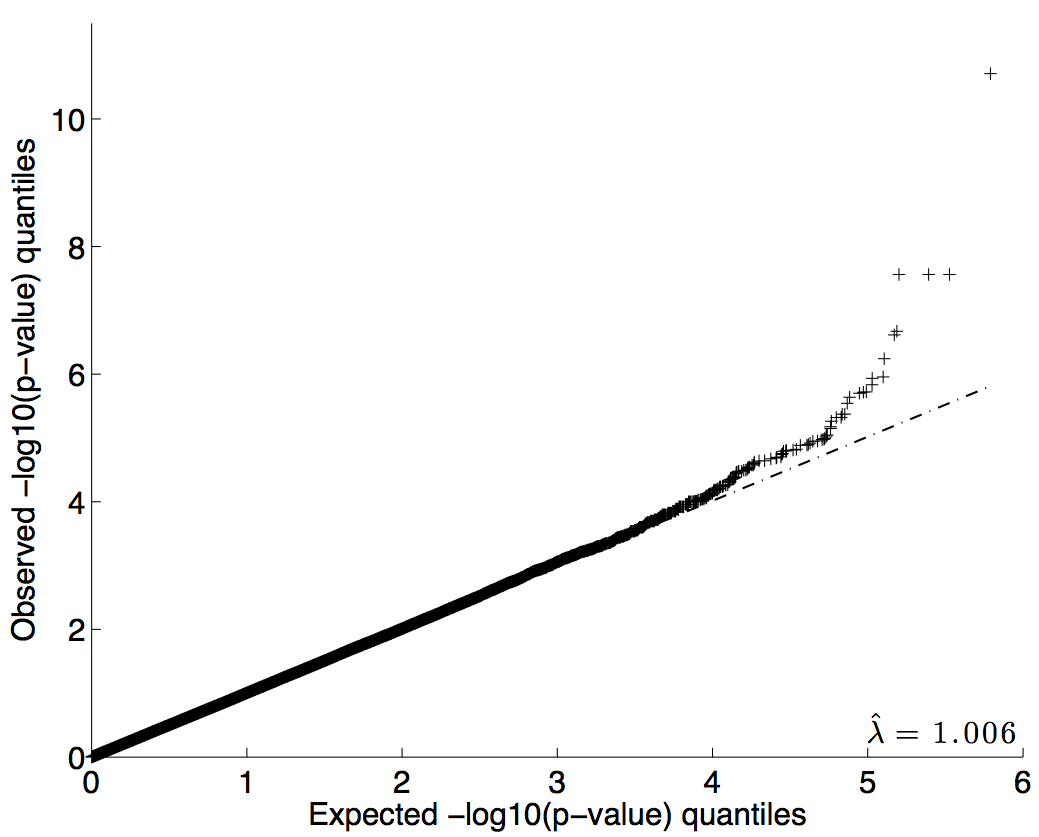}
\caption{The results of \Mendel's HDL$_1$ univariate analysis in the SAFHS data set with global kinship coefficients estimated for all pairs of individuals. Upper: The Manhattan plot graphs roughly one million SNPs against their $-\log_{10}(\text{p-value})$. The horizontal line is the genome-wide significance threshold, $7.3 = -\log_{10}(5 \times 10^{-8})$. Lower: The Q-Q plot graphs the observed $-\log_{10}(\text{p-value})$ quantiles versus their expectations. The genomic control value of $\hat\lambda = 1.006$ derived from this comparison suggests no systematic biases in the data or analysis.}
\label{fig:SAFHS-HDL_1}
\vspace{-0.15in}
\end{figure}

Table~\ref{table:HDL} reports all SNPs with MAF $> 0.01$ that achieve genome-wide significance (p-values less than $5 \times 10^{-8}$) as reported by at least one software package. For the univariate analyses, each software package found the same set of significant SNPs, except that one of \gemma's p-values was slightly short of the significance threshold. Figure~\ref{fig:SAFHS-HDL_1} shows a Manhattan plot and a Q-Q plot from the HDL$_1$ analysis by \Mendel\ given kinship estimates for all pairs of individuals. The results for the other analyses, both univariate and multivariate, were similar. Each \Mendel\ all-pairs univariate analysis had genomic control $\lambda$ in the range 1.002 to 1.006; in default mode, $\lambda$ was in the range 0.992 to 1.022. The various Q-Q plots and associated $\lambda$ values show there is no systematic biases in the data or analysis. In the all-pairs \Mendel\ HDL$_1$ analysis, the grand mean (intercept) was 49.0 $\pm$ 0.8. The SEX covariate was significant in all null models. For example, in the all-pairs \Mendel\ HDL$_\text{Joint}$ analysis with constrained covariates, the SEX effect was 2.4 $\pm$ 0.3 for females and, by design, the opposite for males. The AGE covariate was not significant in any run. For example, again in the all-pairs HDL$_\text{Joint}$ analysis with parameter constraints, the AGE effect was 0.04 $\pm$ 0.02. In the null model for the all-pairs \Mendel\ HDL$_1$ analysis, the additive variance was estimated as 78.8 $\pm$ 9.9, and the environmental variance was estimated as 78.1 $\pm$ 7.2. This gives an overall heritability estimate for HDL$_1$ of 0.50 $\pm$ 0.04. Similar variance estimates were seen in other null models.

For the multivariate analysis without parameter constraints, \Mendel\ is able to include almost twice as many individuals in the analysis as \gemma\ (see Table~\ref{table:run-time}). \gemma\ only includes individuals phenotyped at all component traits and covariates. This probably explains why \Mendel\ finds several more SNPs with significant p-values than \gemma.

\begin{table}[b]
\vspace{-0.10in}
\centering
{\footnotesize
\begin{tabular}{llrrrr}
\toprule
Program  & \multicolumn{2}{l}{Trait \;\;\;\, Analyzed} & Analyzed &  RunTime  &  RAM \\
         &                 \multicolumn{2}{r}{Samples} &    SNPs  & (min:sec) & (GB) \\
\midrule\midrule
\Mendel\ default   &                            & 1357 &  935,392 &    1:51   &  1.2 \\
\Mendel\ all-pairs & HDL$_1$                    & 1357 &  935,392 &    7:49   &  1.2 \\
\fastlmm           &                            & 1397 &  941,546 &   76:11   & 30.0 \\
\gemma             &                            & 1397 &  919,050 &  206:54   &  0.4 \\
\midrule
\Mendel\ default   &                            &  818 &  935,392 &    1:33   &  1.1 \\
\Mendel\ all-pairs & HDL$_2$                    &  818 &  935,392 &    3:25   &  1.1 \\
\fastlmm           &                            &  840 &  934,216 &   49:44   & 18.0 \\
\gemma             &                            &  840 &  914,051 &  180:21   &  0.3 \\
\midrule
\Mendel\ default   &                            &  914 &  935,392 &    1:38   &  1.1 \\
\Mendel\ all-pairs & HDL$_3$                    &  914 &  935,392 &    3:54   &  1.1 \\
\fastlmm           &                            &  939 &  937,208 &   54:58   & 20.0 \\
\gemma             &                            &  939 &  918,626 &  182:26   &  0.3 \\
\midrule
\Mendel\ default   & HDL$_{\text{Joint}}$       & 1388 &  935,392 &    4:08   &  1.2 \\
\Mendel\ all-pairs & {\footnotesize with}       & 1388 &  935,392 &   83:24   &  1.2 \\
\fastlmm           & \multicolumn{2}{l}{{\footnotesize constrained}} & \multicolumn{3}{c}{Not Available}  \\
\gemma             & \multicolumn{2}{l}{{\footnotesize covariates}}  & \multicolumn{3}{c}{Not Available}  \\
\midrule
\Mendel\ default   & HDL$_{\text{Joint}}$       & 1388 &  935,392 &    3:49   &  1.2 \\
\Mendel\ all-pairs & {\footnotesize without}    & 1388 &  935,392 &   80:04   &  1.2 \\
\fastlmm           & \multicolumn{2}{l}{{\footnotesize constrained}} & \multicolumn{3}{c}{Not Available}  \\
\gemma             & {\footnotesize covariates} &  712 &  912,318 &  630:37   &  0.6 \\
\bottomrule
\end{tabular}
}
\vspace{0.05in}
\caption{Comparison of run times and memory (RAM) usage on a typical computer but with adequate RAM to accommodate \fastlmm\ (6 CPU cores at 2.67 GHz, with 48 GB total RAM). The listed run times include reading the data set, performing quality checks, estimating the kinship coefficients, and calculating the association test p-values. All default parameters were used except for minor changes to the quality control thresholds (see text). Also, \Mendel\ was run in both default and all-pairs modes. \Mendel's default mode estimates non-zero global kinship coefficients only for pairs of individuals within the same input pedigree; \Mendel\ in all-pairs mode, \fastlmm, and \gemma\ estimate coefficients for all pairs of individuals. For the multivariate analysis, \Mendel\ includes roughly twice as many individuals as \gemma\ because \gemma\ only analyzes individuals phenotyped at all component traits and covariates. \Mendel\ performs score tests for all SNPs and LRTs for the top SNPs; \fastlmm\ performs LRTs; and \gemma\ by default performs Wald tests, but the user can change this to LRTs or score tests. Using score tests in \gemma\ would make it faster (see text).}
\label{table:run-time}
\vspace{-0.25in}
\end{table}

Table~\ref{table:run-time} tallies the run times and memory footprints from each analysis on a typical personal computer with adequate RAM to accommodate \fastlmm\ (6 CPU cores at 2.67 GHz, with 48 GB total RAM). Even when estimating the global kinship coefficients for all pairs of individuals, each univariate QTL run took \Mendel\ less than 8 minutes to read, quality check, and analyze the data for kinship estimates and association tests, roughly 10\% of the time required for \fastlmm\ and 5\% of the time required by \gemma. (For \gemma, the kinship estimation and association tests are run separately. The run times reported here are their total.)

The three programs use different association test strategies: \Mendel\ performs score tests for all SNPs and LRTs for the top SNPs; \fastlmm\ performs LRTs; and \gemma\ by default performs Wald tests, but the user can change this to LRTs or score tests. For the univariate analyses on a six-core computer, excluding estimation of kinship coefficients, \gemma's run times under the Wald test and LRT options were roughly similar to \fastlmm's; \gemma's run time under the score test option was roughly double \Mendel's in all-pairs mode. This is impressive given \gemma's lack of multithreading. It is kinship estimation, which in practice can be done once per data set, that is substantially slower in \gemma\ (running roughly 135 minutes) than in \fastlmm\ or \Mendel\ (less than 1 minute).

Each trivariate QTL run took \Mendel\ less than 90 minutes. \Mendel\ required roughly one-eighth the time of \gemma\ while analyzing almost twice as many individuals. \Mendel\ is also memory efficient. The univariate and multivariate runs each required less than 1.5 GB of memory, which is well below the amount of RAM in a typical computer. \fastlmm's memory usage is more than 15 times larger than \Mendel's. \gemma\ uses even less memory than \Mendel\ but is considerably slower.

\section{Discussion}

We have implemented an ultra-fast algorithm for QTL analysis of pedigree data or mix of population and pedigree data. In our opinion \Mendel's comprehensive environment for genetic data analysis is a decided advantage. In addition to its exceptional speed and memory efficiency, \Mendel\ can handle multivariate quantitative traits and detect outlier trait values and pedigrees. Most competing programs ignore multivariate traits and outliers altogether. 

A recent review of univariate QTL analysis packages for family data \citep{Cordell2014Review} shows that all the explored packages obtain similar results, leaving speed, features, and ease of use as the important factors in choosing between them. Once the current version of \Mendel\ came out, the authors of the review were kind enough to add a comment (\url{http://www.plosgenetics.org/annotation/listThread.action?root=81847}{www.plosgenetics.org/annotation/listThread.action?root=81847}) to their article observing that \Mendel\ was now the fastest and one of the easiest to use packages they reviewed.

In the SAFHS example data set we used with HDL phenotypes, all the significant SNPs we found had MAF $< 0.01$. Due to these low MAFs, we do not claim these SNPs are strong candidates for further study. However, the key point here is that all four methods found the same SNPs, at least for the univariate analyses. We also note that the p-values are quite similar regardless of whether one uses kinship estimates between all individuals (\Mendel's all-pairs mode) or only between individuals within the same input pedigrees (\Mendel's default mode). This suggests that the input pedigree structures for this data set are substantially correct and complete, with few mistaken or hidden relationships. Obviously, this may not be true for other data sets. By supplying good kinship estimates ignoring pedigree structures, the currently reviewed packages make the hard fieldwork of relationship discovery superfluous.

A future version of \Mendel\ will address its failure to read fractional genotype values. This is simply a logistical issue, as all \Mendel's internal genotype computations are already handled as floating point operations. Another imminent feature is a fourth style of kinship coefficient estimation that allows the user to force theoretical kinship coefficients for pairs of individuals within the same pedigree and estimated kinships for all other pairs.

By supplying a comprehensive, fast, and easy to use package for GWAS on quantitative traits in general pedigrees, we hope to encourage exploitation of family-based data sets for gene mapping. A gene mapping study should collect as large a sample as possible consistent with economic constraints and uniform trait phenotyping. If the sample includes pedigrees, all the better. One should {\em not} let the choice of statistical test determine the data collected; on the contrary, the data should determine the test. Here we have argued that score tests can efficiently handle unrelated individuals, pedigrees, or a mixture of both. For human studies, where controlled breeding is forbidden, nature has provided pedigrees segregating every genetic trait. Many of these pedigrees are known from earlier linkage era studies and should be treasured as valuable resources.

Let us suggest a few directions for future work. The current method works marker by marker and is ill equipped to perform model selection. Lasso penalized regression is available to handle model selection for case-control and random sample data \citep{WuLange08Coordinate,WuChenHastieSobelLange09PenLogistic,Zhou2010-GroupLasso,Zhou2011-WeightedLasso} and can be generalized to variance component models.  Although we have generalized the score test to distributions such as the multivariate $t$, extending it to discrete traits may be out of reach. For likelihood based methods, there simply are no discrete analogues of the Gaussian distribution that lend themselves to graceful evaluation of pedigree likelihoods. Treating case/control data as a 0/1 quantitative variable is a possibility that has been explored by \citet{PirinenDonnellySpencer2013MMM}. The GEE method is another fallback option because it does not depend on precise distributional assumptions.

In rare variant mapping, grouping related SNPs in a variance component may be a good alternative to the mean component models used here. Each variant may be too rare to achieve significance in hypothesis testing. Fortunately, aggregating genotype information within biological units such as genes or pathways offer better power than marginal testing of individual SNPs. See \citet{Asimit2010} for a recent review of aggregation strategies. \citet{Kwee08Kernel} have successfully applied a variance component model for association testing of SNP sets in a sample of unrelated subjects. \citet{Ronnegard2008} consider score tests for random effects models in the context of experimental line crosses. Score tests may well be the key to implementing random effect models in pedigrees.  However, the computational demands are apt to be more formidable than those encountered here with fixed effects models.  In particular, if tests are based simply on local identity-by-descent (IBD) sharing, then the boundaries between pedigrees disappear, and the entire sample collapses to one large pedigree. The required local kinship coefficients can again be well estimated from dense markers, but this demands more computation than the estimation of global kinship coefficients under the mean components model advocated here \citep{Day-Williams2011}. Since inversion of a pedigree covariance matrix scales as the cube of the number of individuals in the pedigree, treating the entire sample as a single pedigree will put a practical upper limit on sample size. There are other issues in implementing variance component models such as assigning p-values and dealing with multivariate traits that are best left to a separate paper.

\section*{Acknowledgement}

\paragraph{Funding:} NIH (GM053275, HG006139, MH059490) and NSF (DMS-1310319).  Burroughs Wellcome Fund Inter-school Training Program in Metabolic Diseases (KHKC).

\bibliography{ScoreTest-arxiv}
\bibliographystyle{apalike}

\end{document}